\documentclass[sigconf, nonacm]{acmart}

\newcommand\vldbdoi{10.14778/3748191.3748195}
\newcommand\vldbpages{3284 - 3298}
\newcommand\vldbvolume{18}
\newcommand\vldbissue{10}
\newcommand\vldbyear{2025}
\newcommand\vldbauthors{\authors}
\newcommand\vldbtitle{\shorttitle} 
\newcommand\vldbavailabilityurl{https://github.com/casys-kaist/DejaVu}
\newcommand\vldbpagestyle{empty} 

\usepackage{package}
\usepackage{macros}

\begin{document}
\pagenumbering{gobble}

\title{\dejavu: Efficient Video-Language Query Engine with Learning-based Inter-Frame Computation Reuse}

\author{Jinwoo Hwang}
\affiliation{%
  \institution{KAIST}
}
\email{jwhwang@casys.kaist.ac.kr}

\author{Daeun Kim}
\affiliation{%
  \institution{KAIST}
}
\email{dekim@casys.kaist.ac.kr}

\author{Sangyeop Lee}
\affiliation{%
  \institution{KAIST}
}
\email{sangyeop-lee@casys.kaist.ac.kr}

\author{Yoonsung Kim}
\affiliation{%
  \institution{KAIST}
}
\email{yskim@casys.kaist.ac.kr}

\author{Guseul Heo}
\affiliation{%
  \institution{KAIST}
}
\email{gsheo@casys.kaist.ac.kr}

\author{Hojoon Kim}
\affiliation{%
  \institution{KAIST}
}
\email{hojoonkim@casys.kaist.ac.kr}

\author{Yunseok Jeong}
\affiliation{%
  \institution{KAIST}
}
\email{ysjeong@casys.kaist.ac.kr}

\author{Tadiwos Meaza}
\affiliation{%
  \institution{KAIST}
}
\email{tadiwos@casys.kaist.ac.kr}

\author{Eunhyeok Park}
\affiliation{%
  \institution{POSTECH}
}
\email{eh.park@postech.ac.kr}

\author{Jeongseob Ahn}
\affiliation{%
  \institution{Korea University}
}
\email{jsahn@korea.ac.kr}

\author{Jongse Park}
\affiliation{%
  \institution{KAIST}
}
\email{jspark@casys.kaist.ac.kr}

\begin{abstract}
Recently, Video-Language Models (VideoLMs) have demonstrated remarkable capabilities, offering significant potential for flexible and powerful video query systems.
These models typically rely on Vision Transformers (ViTs), which process video frames individually to extract visual embeddings.
However, generating embeddings for large-scale videos requires ViT inferencing across numerous frames, posing a major hurdle to real-world deployment and necessitating solutions for integration into scalable video data management systems.
This paper introduces \dejavu, a video-language query engine that accelerates ViT-based VideoLMs by reusing computations across consecutive frames. 
At its core is \reusevit, a modified ViT model specifically designed for VideoLM tasks, which learns to detect inter-frame reuse opportunities, striking an effective balance between accuracy and reuse.
Although \reusevit significantly reduces computation, these savings do not directly translate into performance gains on GPUs. 
To overcome this, \dejavu integrates memory-compute joint compaction techniques that convert the FLOP savings into tangible performance gains. 
Evaluations on three VideoLM tasks show that \dejavu accelerates embedding generation by up to a 2.64$\times$ within a 2\% error bound, dramatically enhancing the practicality of VideoLMs for large-scale video analytics.
\end{abstract}

%

\maketitle

\pagestyle{\vldbpagestyle}
\begingroup\small\noindent\raggedright\textbf{PVLDB Reference Format:}\\
\vldbauthors. \vldbtitle. PVLDB, \vldbvolume(\vldbissue): \vldbpages, \vldbyear.\\
\href{https://doi.org/\vldbdoi}{doi:\vldbdoi}
\endgroup
\begingroup
\renewcommand\thefootnote{}\footnote{\noindent
This work is licensed under the Creative Commons BY-NC-ND 4.0 International License. Visit \url{https://creativecommons.org/licenses/by-nc-nd/4.0/} to view a copy of this license. For any use beyond those covered by this license, obtain permission by emailing \href{mailto:info@vldb.org}{info@vldb.org}. Copyright is held by the owner/author(s). Publication rights licensed to the VLDB Endowment. \\
\raggedright Proceedings of the VLDB Endowment, Vol. \vldbvolume, No. \vldbissue\ %
ISSN 2150-8097. \\
\href{https://doi.org/\vldbdoi}{doi:\vldbdoi} \\
}\addtocounter{footnote}{-1}\endgroup

\ifdefempty{\vldbavailabilityurl}{}{
\vspace{.3cm}
\begingroup\small\noindent\raggedright\textbf{PVLDB Artifact Availability:}\\
The source code, data, and/or other artifacts have been made available at \url{\vldbavailabilityurl}.
\endgroup
}

\section{Introduction}
Video data is pervasive in today's world, playing a critical role in various applications such as autonomous machines~\cite{deb2018pedestrians,gawande2020pedestrian,hadidi2018distributed}, education~\cite{largeedu}, healthcare~\cite{fcntransformer}, security and surveillance~\cite{isee, biometric}, e-commerce~\cite{poet, attract}, and many others~\cite{marioqa, hero}. 
Recently, Video-Language Models (VideoLMs) have demonstrated remarkable capabilities, offering significant potential for flexible and powerful video analytics tasks~\cite{tutorialvqa, medvidqa, lavila, prompting, justask,frozenbilm}.
While these powerful algorithmic advances promise enormous potential, the massive computational demand poses a significant hurdle for the widespread adoption of VideoLMs for video query systems.

The excessive computational demand arises from two primary factors.
First, modern VideoLMs heavily rely on Vision Transformers (ViTs)~\cite{vit} as their core engine for visual feature extraction.
ViTs contain hundreds of millions of parameters and require tens of billions of FLOPs for each inference.
Second, ViT inference must be performed iteratively across numerous video frames, which can be exceptionally large in number (e.g., a 1-hour video sampled at 2 FPS contains 7,200 frames).
Consequently, utilizing VideoLMs for large-scale video analytics becomes nearly infeasible, even with multi-GPU datacenter resources.

\begin{figure}
    \centering
    \includegraphics[width=0.85\linewidth]{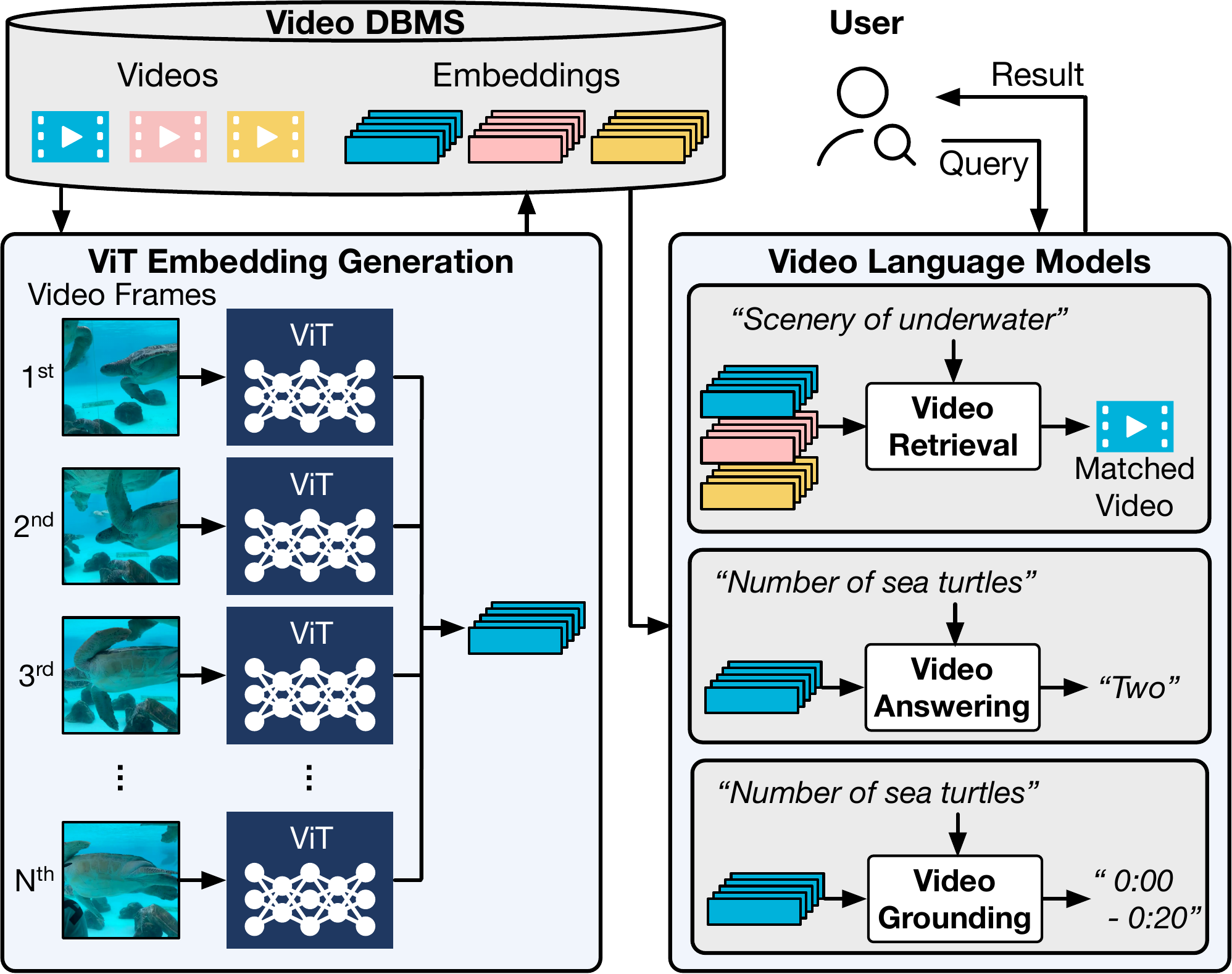}
    \caption{Overview of video language query systems supporting three example VideoLMs. Modern VideoLMs employ vision transformer (ViT) as their core engine.}
    \label{fig:video-understanding-overview}
\end{figure}
Figure~\ref{fig:video-understanding-overview} provides an overview of video-language query systems, where ViTs process video frames to extract visual embeddings, which are then fed into query-specific operations such as video retrieval, video question answering, and video question grounding.
The critical role of ViTs in this process emphasizes the importance of improving ViT inference efficiency to unleash the potential of VideoLMs for building video-language query systems.

Predictably, the vital need for efficient ViT inferencing has led to a large body of acceleration work in recent years~\cite{dynamicvit, ats, evit, spvit, diffrate, pumer, leopard, spatten, heatvit, fact, tome, tokenpooling}.
Most of these works focus on \textit{intra}-frame computation reuse, targeting inter-token feature similarities since they aim to speed up a single ViT inference run processing one image input.
While effective for single images, these methods overlook additional opportunities present in video data that exhibit \textit{inter}-frame similarities across multiple consecutive frames.
Recently, several works~\cite{cmc,eventful,vid-tldr} have pioneered the exploitation of \textit{inter}-frame computation reuse for ViT inferencing on video applications.
However, these works have the following two limitations, which leave them in a position of questionable utility for most practical use cases: 
\begin{itemize}[itemsep=0pt, parsep=0pt, topsep=1pt, partopsep=1pt,left=5pt..2em]
\item \textbf{Limitation 1: } These methods manually identify reuse opportunities for given models and fixedly enable computation reuse, making it considerably difficult to locate the sweet spot on the accuracy-reuse tradeoff space. 
\item \textbf{Limitation 2: } Although these techniques enable significant FLOP reduction, they do not yield corresponding performance gains, as the remaining computations are sparse and, therefore, inefficient on GPUs.
\end{itemize}
To address these limitations, this work proposes \dejavu, an algorithm-system co-designed query engine that leverages a learning-based approach to automatically identify the computation reuse opportunities and find the sweet spot in the tradeoff space between accuracy and reuse.

The key challenges are (1) to maximize the reuse opportunities without causing significant accuracy loss in VideoLM applications and (2) to effectively translate the FLOPs savings into performance gains.
\dejavu tackles these challenges by making the following two key contributions, each corresponding to the two limitations:
\begin{description}[labelindent=0.3em,nolistsep,leftmargin=1.0em]
\item[(1)]
\textbf{\emph{\reusevit: Reuse-enabling ViT model.}}
We develop a modified ViT model, referred to as \reusevit, that reuses precomputed values for similar tokens between frames.
To maximize reuse opportunities, we devise a frame referencing strategy that determines inter-frame reference chains for computation reuse.
\reusevit contains \emph{Decision Layers} that dynamically determine when to reuse computations by processing multiple inputs such as cosine similarity, attention scores, codec metadata, and reference types, allowing it to learn the importance and interactions of these hints automatically.
Additionally, it employs \emph{Restoration Layers} that learn to calibrate the reused computation values based on the changes that occurred in the current frame.
\quad To train \reusevit, we use reparameterization with soft gating via the Gumbel-Softmax mechanism, enabling backpropagation through discrete decisions.
This allows the model to mimic hard gating, allowing gradients to flow. 
Further, we model error propagation by training with grouped frames to mitigate error accumulation from reusing computations across multiple frames.
\item[(2)]
\textbf{\emph{Memory and compute compactions for fast and efficient \reusevit inferencing.}}
To achieve significant performance gains from these algorithmic innovations, we propose three key system-level techniques.
(1) We introduce \textit{layer-wise computation scheduling}, which processes computations for multiple frames in a layer-by-layer manner. 
(2) This approach enables \textit{cached memory compaction}, a memory management technique that clears caches after each layer is completed, allowing for optimized memory usage. 
Consequently, this increases the batch size during inference, leading to better hardware utilization.
(3) Additionally, we implement \textit{sparse computation compaction}, which restructures irregular data patterns caused by computation reuse into dense computations suitable for efficient GPU execution. 
By consolidating tokens from multiple frames, we create more regular matrices, improving the efficiency of matrix multiplication operations on GPUs, particularly when reuse rates are high.
\end{description}
To demonstrate the effectiveness of \dejavu, we perform evaluations using three different VideoLMs -- (1) Video Retrieval, (2) Video Question Answering, and (3) Video Question Grounding -- on their respective datasets. 
We observe that \dejavu achieves throughput improvement of 1.81$\times$, 2.64$\times$, and 2.54$\times$ for the three tasks, respectively, within 2\% error bound. 
Under the same accuracy constraints, the state-of-the-art systems using inter-frame computation reuse offer only 1.32$\times$,  2.08$\times$, 2.20$\times$ speedup for the respective tasks.

These significant performance gains, coupled with minimal accuracy loss, underscore the solution's potential to bridge the computational gap, paving the way for VideoLMs to unlock new capabilities.
\section{Background and Motivation}
\subsection{Video Query Processing}
\label{sec:video_indexing}
A rich body of research explores accelerating large-scale video queries using AI models.
We categorize them into three groups.

\niparagraph{Task-specific CNN pipelines.}
Early approaches~\cite{noscope, blazeit, everest:sigmod:2022, figo:sigmod:2022} accelerate queries by training small, query-specific models as approximations for computationally expensive deep learning pipelines.
Systems such as NoScope~\cite{noscope} and BlazeIt~\cite{blazeit} train lightweight classifiers or regression models tailored to each query, significantly reducing inference costs for known object classes.
However, these methods are inherently inflexible, as they require model training or selection for every new query, making them unsuitable for quickly adapting to new or changing requirements.

\niparagraph{Task-agnostic proxy embeddings.}
The second category~\cite{tasti:vldb:2021, seiden:vldb:2023} precomputes a single embedding for video frames to allow fast similarity-based retrieval without query-specific training.
Methods such as TASTI~\cite{tasti:vldb:2021} cluster frames in a learned embedding space, allowing queries to be answered using indexed embeddings rather than full model inference.
%
While these approaches eliminate the per-query model burden, they may struggle with open-ended queries (i.e., broad or natural-language requests) if novel or uncommon concepts are poorly represented in precomputed embeddings.

\niparagraph{Vision-Language Pretrained (VLP) embeddings.}
More recent approaches~\cite{zelda:2023,seesaw:pacmmod:2023} provide a more general solution by leveraging VLP models like CLIP~\cite{clip}, which map image and text queries into a shared semantic space.
These models support open-ended, natural-language-based video queries without requiring predefined object classes or embedding indexes.

However, recent works~\cite{seesaw:pacmmod:2023, sketchql:pacmmod:2024} illustrate that relying solely on models has limitations in practical video retrieval scenarios.
While VLP models exhibit strong generalization capabilities, VLP models often struggle with uncommon or domain-specific queries where training data is sparse.
To address these concerns, \dejavu leverages VideoLMs, advanced models built on top of VLP architectures that specialize in understanding video context.

\subsection{Video-Language Models (VideoLMs)} 
VideoLMs extend VLP architectures to support complex tasks including video retrieval, video question answering, and video question grounding.
For instance, video retrieval involves matching a textual description (e.g., "Show me a clip of scenery of underwater") to the most relevant video in a large corpus.
Video question answering asks natural-language questions about the visual content, such as “How many sea turtles appear?”, requiring the model to understand both objects and context.

\niparagraph{Computational challenges.}
Despite their impressive capabilities, the high computational complexity of VideoLMs remains a critical hurdle for potential applications.
A major contributor to VideoLM overhead is the embedding generation process using large-scale VLP models~\cite{florence, flava, clip, vilt, alpro, blip, align, albef, lfvila, klite}, which commonly rely on Vision Transformer (ViT)~\cite{vit} architectures.
ViTs often comprise hundreds of millions of parameters and require billions of FLOPs per inference which need to run on numerous video frames.
\begin{figure}
    \centering
    \includegraphics[width=0.95\linewidth]{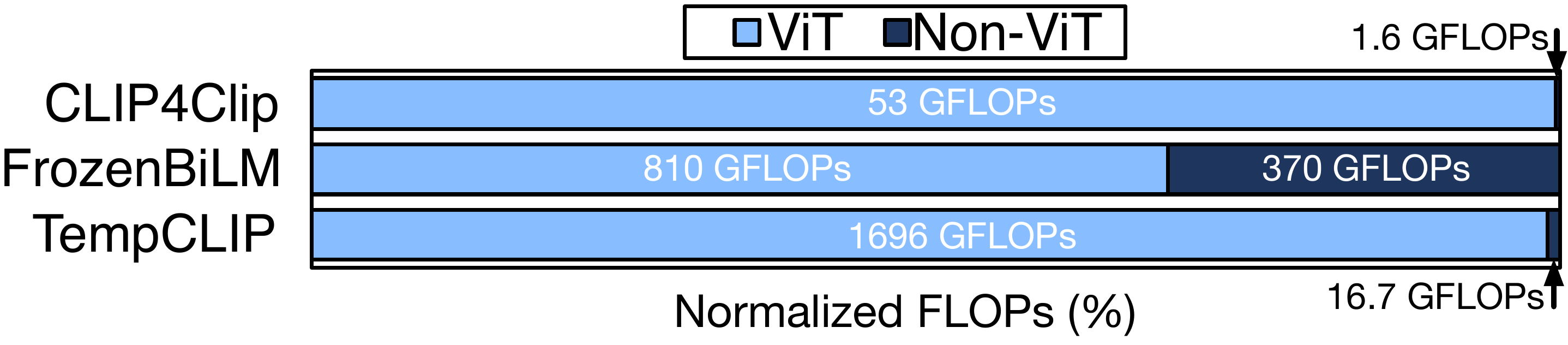}
    \caption{FLOPs breakdown across three VideoLM tasks: video retrieval (CLIP4Clip), video question answering (FrozenBiLM), and video question grounding (TempCLIP).}
    \label{fig:flops-video-model}
\end{figure}
Figure~\ref{fig:flops-video-model} illustrates the computational breakdown for three VideoLMs.
Notably, embedding generation via ViT dominates the overall FLOPs, underscoring the need for efficient ViT acceleration.
\subsection{Vision Transformer (ViT) Acceleration}
\label{sec:vit}
\niparagraph{Architecture of ViT.}
\begin{figure}
    \centering
    \includegraphics[width=0.95\linewidth]{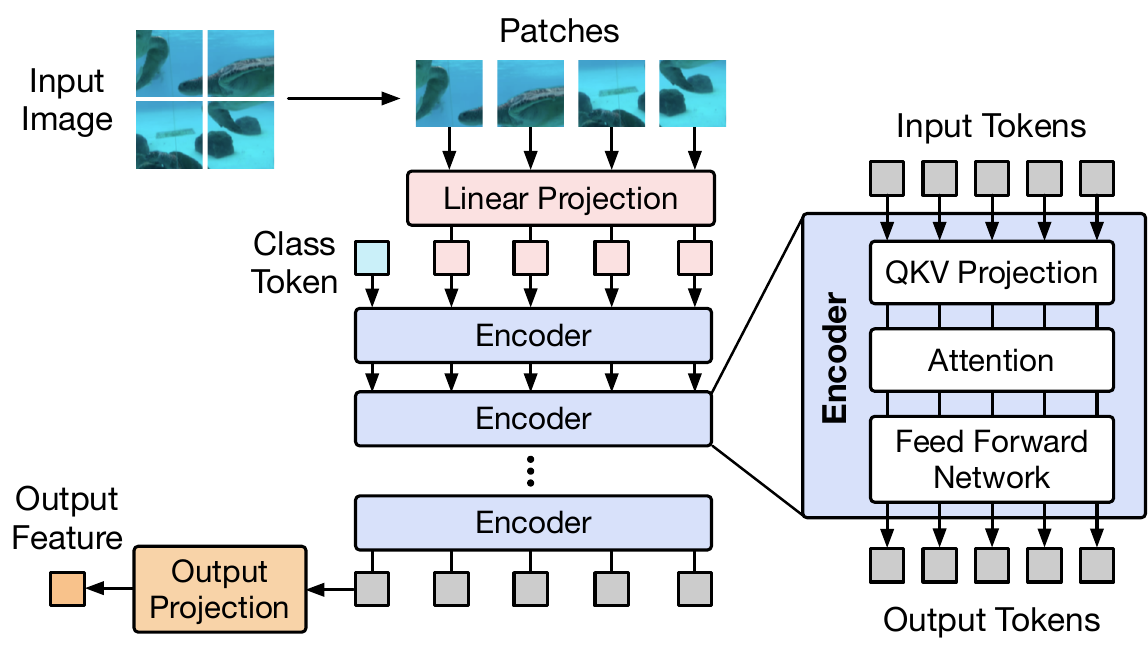}
    \caption{ViT model architecture.}
    \label{fig:vit}
\end{figure}
Figure~\ref{fig:vit} illustrates ViT architecture, which adapts the transformer architecture from natural language processing to the vision domain.
In ViTs, an input image or video frame is partitioned into fixed-size patches, which are linearly embedded to form a sequence of tokens.
A class token (CLS) is appended to this sequence to aggregate global information.
The tokens pass through Transformer encoder layers, performing query-key-value projections, multi-head self-attention, and feed-forward network.
%
%

\niparagraph{Existing acceleration methods.}
To address the computational demands of ViTs, various methods exploit redundancies in the data.
Acceleration techniques focusing on inter-token redundancies reduce computation by pruning less important tokens or merging similar ones within a single image~\cite{dynamicvit, ats, evit, spvit, pumer, leopard, spatten, heatvit, fact, tome, tokenpooling, diffrate}.
While these pruning and merging methods effectively skip unnecessary computations within a single image or frame, they ignore the redundancy across frames.
In large-scale video analytics, small frame-to-frame changes can be exploited to a much greater extent.
Hence, single-image token pruning might still repeat most computations for each of these frames, making it less effective for long videos in which large swaths of patches remain nearly identical over short time intervals.

More recently, other methods have attempted to leverage inter-frame computation reuse in video data~\cite{eventful,cmc,vid-tldr}, offering greater computational savings.
%
Among them, CMC~\cite{cmc} and Eventful Transformer~\cite{eventful} are most relevant to \dejavu, as they explicitly reuse partial computations across frames.
In contrast, vid-TLDR~\cite{vid-tldr} merges tokens temporally, reducing redundancy through token aggregation rather than direct computation reuse.
We provide detailed comparisons against CMC and Eventful Transformer in Section~\ref{sec:methodology}.

\subsection{Limitations of Video ViT Acceleration}
\label{subsec:prior-works-limitation}
While promising, video-targeted acceleration methods exhibit limitations that hinder their practical utility.

\niparagraph{Challenges in balancing reuse and accuracy.}
Existing methods rely on manually designed strategies for computation reuse, which can make it substantially difficult to locate the optimal balance between accuracy and computational savings.
For instance, Eventful Transformer~\cite{eventful} requires fixing the number of tokens that reuse computation at each layer.
Due to the large search space created by the tens of encoder layers and the variability in video content, it is challenging to predict the consequences of increasing reuse in certain layers.
As a result, these methods may lead to a suboptimal tradeoff between computation reuse and accuracy.

\niparagraph{Challenges in realizing FLOP savings as speedups.}
While reducing computational complexity (FLOPs) is crucial, achieving actual speedups requires addressing runtime factors. 
Mixing computed and reused tokens leads to sparse computations, causing inefficiencies on GPUs optimized for dense workloads.
Our empirical analysis shows that existing video-targeted ViT acceleration techniques~\cite{cmc, eventful} deliver limited speedups due to overheads in memory usage, data movement, and hardware utilization (see Section~\ref{sec:evaluation}).
Furthermore, some prior acceleration works~\cite{cmc} rely on specialized hardware accelerators, which are not readily available in standard commodity systems, limiting the accessibility.

\finding{
These limitations motivate us to design a customized ViT model, dubbed \reusevit, which can automatically identify computation reuse opportunities in video data, while carefully balancing the accuracy-reuse tradeoff.
Additionally, we introduce memory-compute joint compaction techniques to effectively convert \reusevit's FLOP savings into tangible performance gains.
%
}
\section{Model Architecture of \reusevit}
\label{sec:design-simvit}
To address computational challenges in ViT-based VLMs, we propose \reusevit, a model that automatically identifies safe (accuracy-preserving) computation reuse opportunities to accelerate inference.
\reusevit is designed to maximally reduce redundant computations while maintaining high accuracy by leveraging inter-frame similarities in video data.
%
%
\subsection{Frame Selection for Computation Reuse}
\label{subsec:computation-ordering}
Determining which frames to reference is critical to maximizing computation reuse opportunities.
In video data, especially at low frame rates common in VideoLM tasks, frame contents can drift significantly over time.
%
%
Hence, \reusevit reorders frames to allow referencing both the past and future frames, potentially boosting the likelihood of finding similar content.

\niparagraph{Sequential frame computation.}
\begin{figure}[t]
    \centering
    \includegraphics[width=\linewidth]{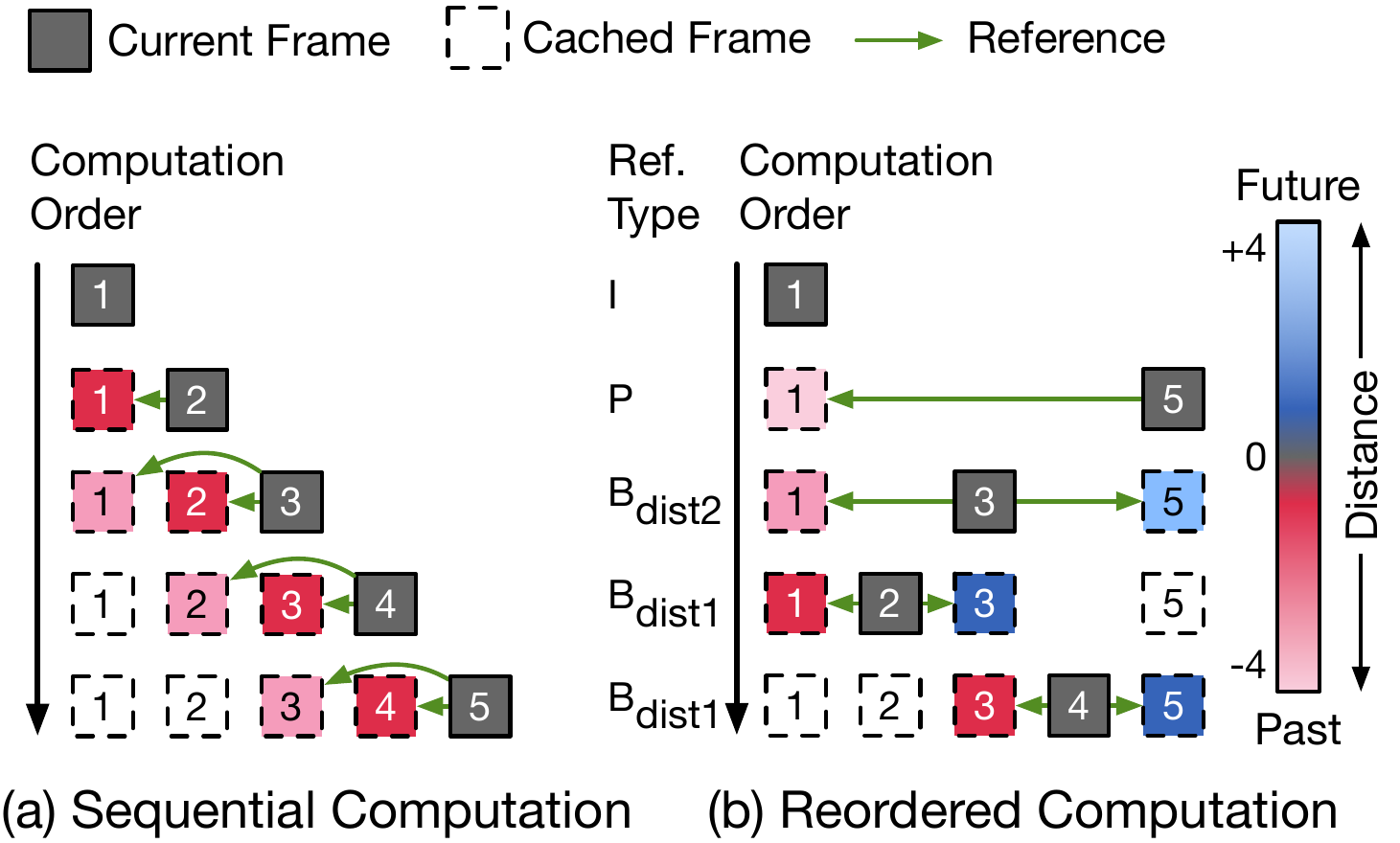}
    \caption{(a) A basic approach where each frame references preceding frames sequentially, (b) Our proposed reordering where frames reference both past and future frames.}
    \label{fig:frame-reordering}
\end{figure}
Figure~\ref{fig:frame-reordering}(a) illustrates a basic strategy where each new frame references one or more preceding frames. 
Because nearer frames generally exhibit higher temporal similarity, referencing them provides reuse benefits.
However, including frames further back tends to yield diminishing returns due to overlapping patches that offer little additional information.

\niparagraph{Reordered frame computation.}
To harness both past and future frames, \reusevit processes frames in an out-of-order fashion, as shown in Figure~\ref{fig:frame-reordering}(b).
We categorize frames into four types, I-frames (computed independently), P-frames (referencing a previous frame), B\textsubscript{dist2}-frames (referencing frames two steps away), and B\textsubscript{dist1}-frames (referencing immediate neighbors), reflecting terminology akin to video codecs.
Following a pattern of I $\rightarrow (P \rightarrow B\textsubscript{dist2} \rightarrow B\textsubscript{dist1} \rightarrow B\textsubscript{dist1}) \rightarrow … $ helps capture bidirectional temporal redundancies that purely sequential schemes may overlook.
By referencing both directions, we increase the potential for reuse and reduce the frequency of full computations.
When choosing references, \reusevit evaluates the quality and temporal distance of candidate frames to preserve accuracy without incurring excessive overhead.
Subsequent sections detail this decision-making process.
\subsection{Layer Selection for Reuse}
%
Deciding which layers within the ViT architecture are suitable for computation reuse is crucial for maximizing efficiency gains without compromising performance.
%

\niparagraph{FLOPs breakdown of ViT layers.}
\begin{figure}[t]
    \centering
    \includegraphics[width=\linewidth]{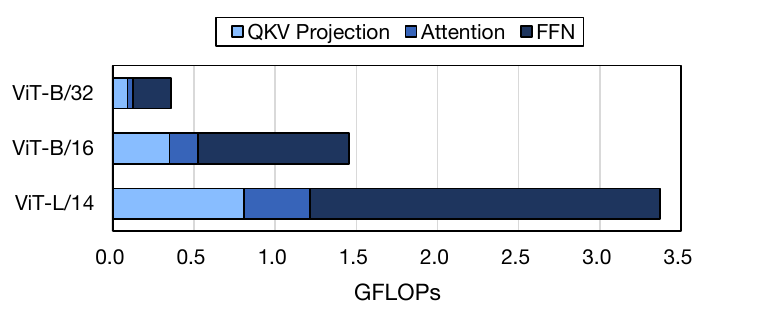}
    \caption{FLOPs breakdown of core computations within a single encoder layer of vision transformers at different scales.}
    \label{fig:ViT-FLOPs}
\end{figure}
\begin{figure*}
\centering
\includegraphics[width=\linewidth]{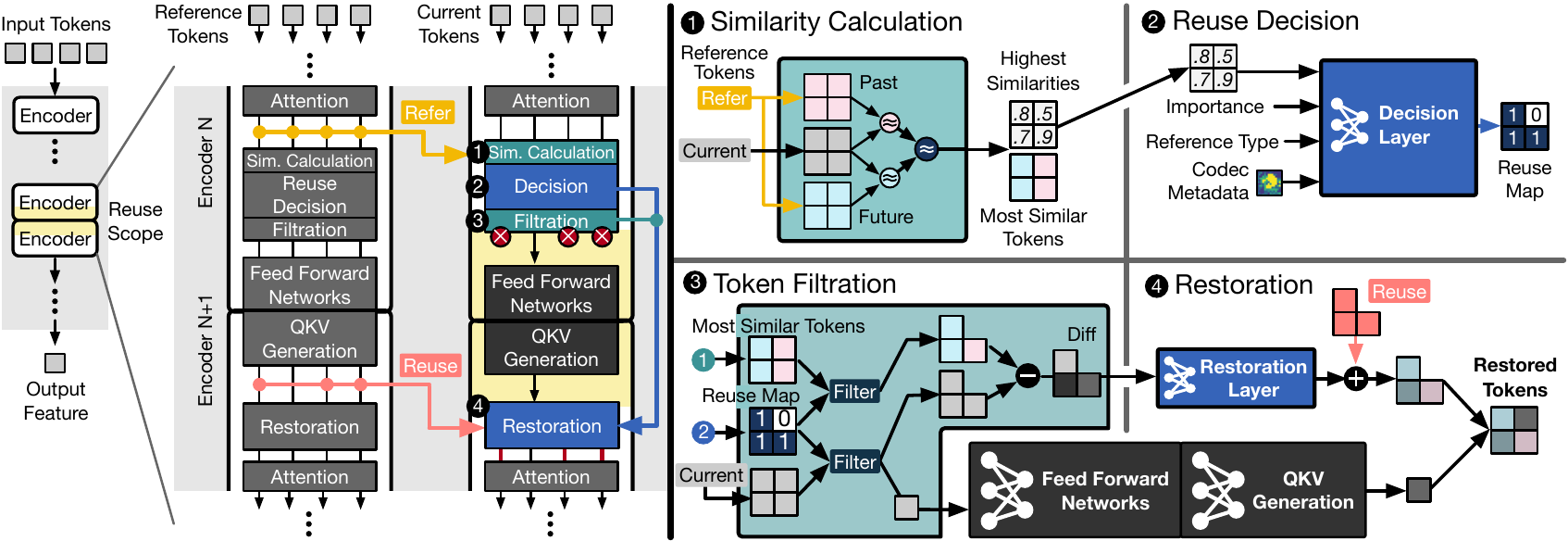}
\caption{Overview of \simvit model operations.}
\label{fig:overview}
\end{figure*}
Figure~\ref{fig:ViT-FLOPs} presents the FLOPs breakdown analysis results for a transformer encoder layer of three different ViT variants.
The results suggest that the query-key-value (QKV) projection and the feed-forward network (FFN) are the primary consumers of computational resources in ViTs.
Unlike large language models (LLMs), where the self-attention layer is the main bottleneck due to longer sequence lengths, ViTs process shorter sequences (around 256 patches). 
Thus, in ViTs, the self-attention layer contributes much less to the overall computational cost relative to the QKV projection and FFN layers.

\niparagraph{Computation patterns of layers.}
Besides computational cost, we consider computation patterns of each layer.
The QKV projection and FFN operations are applied independently to each token, without inter-token dependencies, suiting them for computation reuse.
In contrast, the self-attention operation involves interactions among all tokens, making reuse more challenging and potentially affecting model output.
Given these considerations, we focus on reusing computations in the QKV projection and FFN layers, which are the most computationally intensive and token-independent.
%

%
\subsection{Criteria for Computations Reuse} 
\label{subsec:simvit-model}
An essential aspect of \reusevit is determining when to reuse computations for specific tokens.
We introduce two key components: (1) a \emph{Decision Layer} that decides whether to reuse computations, and (2) a \emph{Restoration Layer} that calibrates reused computations to align with the current frame’s context.
Figure~\ref{fig:overview} depicts the overall architecture of \reusevit.

\niparagraph{Decision layer.}
The decision layer identifies tokens suitable for computation reuse by assessing multiple informative cues for each token.
By integrating various inputs, the layer makes nuanced decisions without relying on hand-crafted rules.
The decision layer takes as input a concatenation of the following features:
\begin{itemize} [labelindent=0.3em,leftmargin=1.0em]
\item \textbf{Similarity Measure ($s$):}
We compute the cosine similarity between the current token and corresponding tokens in the reference frames.
A higher similarity suggests that the token content has remained unchanged, making it a candidate for reuse:
\begin{equation}
s_i = \max\left( \cos\left( T^{\text{cur}}_i, T^{\text{past}}_i \right),\ \cos\left( T^{\text{cur}}_i, T^{\text{future}}_i \right) \right)
\end{equation}
Here, $T^{\text{cur}}_i \in \mathbb{R}^{D}$ represents the $i$-th token of the current frame, while $T^{\text{past}}_i$ and $T^{\text{future}}_i$ are the corresponding tokens from the previous and future reference frames.
This strategy follows prior work on token merging using similarity metrics~\cite{tome, heatvit, diffrate}.
\item \textbf{Token Importance ($t$):}
We use the attention weights from the class token to estimate each token’s importance.
Tokens with higher attention are more critical and may require fresh computation.
This method is consistent with existing approaches that prune tokens based on class-token attention~\cite{ats, evit, evovit, diffrate}.
\item \textbf{Reference Type ($r$):}
Reference frame type (e.g., I-frame, P-frame, B\textsubscript{dist1}, B\textsubscript{dist2}) offers insight into the reference's temporal proximity, aiding assessment of reused computation reliability
%
\item \textbf{Codec Metadata ($c$)}:
Metadata from video codecs offers block-wise hints about spatio-temporal redundancies, providing insights into areas where video content undergoes motion or structural changes.
These metadata signals, such as motion vectors and residuals, have been leveraged in prior research~\cite{cova,packetgame,crucio} to guide computational optimizations.
\end{itemize}
By combining these inputs, the decision layer, implemented as a simple two-layer MLP, makes informed decisions.
\begin{align}
v_i &= \text{concat}(s_i, t_i, r_i, c_i), \quad i = 1, \ldots, N \\
d_i &= \text{MLP}_{\text{decision}}(v_i) \\
\mathcal{M}_i &= \begin{cases}
1, & \text{if } d_i > 0 \\
0, & \text{otherwise}
\end{cases}, \quad i = 1, \ldots, N
\label{eq:hard-gating}
\end{align}
where $\mathcal{M}_i \in {0,1}$ indicates whether to reuse ($1$) or recompute ($0$) the computations for token $i$.
%
%
By combining similarity scores with token importance and codec metadata, the decision layer captures both visual and semantic cues, striking a balance between accuracy and reuse.
This data-driven approach enables the model to adapt effectively to diverse video scenarios without manual tuning of thresholds or heuristics.

\niparagraph{Token filtration.}
Based on the \bluetext{reuse map} $\mathcal{M}$, tokens are partitioned into recompute tokens $C$ and reuse tokens $R$.
\begin{align}
C &= { T^{\text{cur}}_i \mid \mathcal{M}_i = 0 }, \quad \text{(re\underline{c}ompute tokens)} \\
R &= { T^{\text{cur}}_i \mid \mathcal{M}_i = 1 }, \quad \text{(\underline{r}euse tokens)}
\end{align}
Recompute tokens $C$ undergo the standard feed forward network and query-key-value projection.
\begin{equation}
\tilde{C}^{\text{cur}} = \text{QKV} \left( \text{FFN} \left( C^{\text{cur}} \right) \right)
\end{equation}

\niparagraph{Restoration layer.}
To adjust for discrepancies between reused computations and the current frame, we introduce a restoration layer that calibrates the reused tokens.
For each reuse token $i$, we compute the difference between the tokens.
\begin{equation}
\Delta R_i = R^{\text{cur}}_i - R^{\text{ref}}_i
\end{equation}
This difference captures changes between the current and reference tokens.
The restoration layer processes $\Delta R_i$ using a small two-layer MLP with a hidden size of 128, which is significantly smaller than the hidden size of 1,024 used in the FFN layers.
This design choice reduces computational overhead while efficiently obtaining a calibration value for the change in the input.
\begin{equation}
\hat{R}^{\text{cur}}_i = \tilde{R}^{\text{ref}}_i +
\text{MLP}_{\text{restoration}}( \Delta R_i ) \end{equation}
where $\tilde{R}^{\text{ref}}_i$ is the reused computation from the reference frame, and $\hat{R}^{\text{cur}}_i$ is the calibrated token for the current frame.
This calibration effectively improves accuracy with minimal computational overhead (4\% compared to standard QKV and FFN computation).

\niparagraph{Token Reconstruction}
We reconstruct the full set of tokens $\hat{T}^{\text{cur}}$ by combining the recomputed tokens $\tilde{C}^{\text{cur}}$ and the reused tokens $\hat{R}^{\text{cur}}$ according to the \bluetext{reuse map} $\mathcal{M}$.
\begin{equation}
\hat{T}^{\text{cur}}_i = \begin{cases}
\tilde{C}^{\text{cur}}_{j}, & \text{if } \mathcal{M}_i = 0 \\
\hat{R}^{\text{cur}}_{k}, & \text{if } \mathcal{M}_i = 1
\end{cases}
\end{equation}
where $i$ indexes the original token sequence, and $j$ and $k$ index the recomputed and calibrated reused tokens, respectively.
This process maintains the original order of tokens, preserving positional relationships within the model.

%
%
The low-complexity design of the decision and restoration modules (e.g., small MLPs) ensures modest overhead relative to full inference, yielding significant net savings in large-scale deployments despite the small cost of restoring reused tokens.

\section{Training \reusevit}
\label{sec:decision-making}
\reusevit's efficiency requires training its decision and restoration layers for a given ViT model and dataset, keeping the pre-trained ViT model frozen.
This section outlines these training techniques, focusing on handling discrete reuse decisions and modeling error accumulation through grouped frame training.
%
\subsection{Handling Discrete Reuse Decisions}
\label{sec:backpropagation}
A key training challenge for \reusevit is handling discrete reuse decisions, which obstruct gradient flow through the gating mechanism during backpropagation.
%
Specifically, the binary decisions to reuse or recompute computations for tokens hinder gradient-based optimization essential for training.

To enable gradient flow through the gating mechanism, we approximate the hard binary decisions with continuous values during training.
This soft gating mechanism allows the decision layer to be trained end-to-end using backpropagation without the need for specialized techniques.
By replacing the hard decisions in Equation~\ref{eq:hard-gating} with continuous approximations, we ensure that gradients can propagate through the layer, facilitating effective learning.
\begin{align}
\mathcal{M}_{\text{soft}} &= \text{GumbelSoftmax}\left( \text{MLP}_{\text{decision}}(v) \right) \approx [0, 1]^{N} \label{eq:gumbel} \\
\hat{T}_{\text{soft}}^{\text{cur}} &= \mathcal{M}_{\text{soft}} \odot \hat{T}^{\text{cur}} + \left( 1 - \mathcal{M}_{\text{soft}} \right) \odot \tilde{T}^{\text{cur}} \label{eq:hat-t-soft}
\end{align}
Here, $\odot$ denotes element-wise multiplication, $\mathcal{M}_{\text{soft}}$ is the soft mask for all $N$ tokens, $\hat{T}^{\text{ref}}$ represents the reused computations from reference frames, and $\tilde{T}^{\text{cur}}$ represents the recomputed tokens of the current frame.
We found that gradually lowering the Gumbel-Softmax temperature over the course of training helps the model transition smoothly from a fully soft gating regime toward more selective token reuse.
This annealing avoids sudden spikes in the reuse mask while preserving stable gradient flow.

At inference time, we apply the hard decisions to realize the computational savings achieved by \reusevit.
\subsection{Loss Function for Accuracy and Efficiency}
\label{sec:balanced-loss}
%
%
To train \reusevit effectively, we formulate a loss function balancing model accuracy and computational efficiency, aiming for performance gains without degradation in prediction quality.

\niparagraph{Training objective.}
Our objective is to ensure that the approximated features produced by \reusevit remain close to the original features generated by the unmodified ViT model.
By preserving the original embeddings, we maintain the model’s predictive performance across different tasks without the need for task-specific fine-tuning.
This self-supervised approach allows the model to generalize effectively in practical systems.

\niparagraph{Similarity loss.}
We use the cosine similarity metric to encourage the approximated features to remain close to the original features of the VLP model.
\begin{align}
\mathcal{L}_{\text{sim}} &= 1 - \cos\left( Z^{\text{cur}}, \hat{Z}^{\text{cur}} \right) \label{eq:sim_loss}
\end{align}
Here, $Z^{\text{cur}}$ is the original final feature vector, and $\hat{Z}^{\text{cur}}$ is the corresponding feature vector after computation reuse and calibration.
However, using only the similarity loss may discourage computation reuse, as the model can minimize loss by recomputing all tokens, negating efficiency gains.

\niparagraph{Reuse loss.}
To incentivize reuse, we introduce a reuse loss.
\begin{align}
\mathcal{L}_{\text{reuse}} &= \frac{1}{L N} \sum_{l=1}^{L} \sum_{i=1}^{N} \mathcal{M}_{l,i}
\label{eq:reuse_loss}
\end{align}
This loss term represents the average reuse rate across all tokens and layers. By maximizing $\mathcal{L}_{\text{reuse}}$, we encourage the model to reuse more computations, promoting computational efficiency.

\niparagraph{Combined loss function.}
The overall loss function combines the similarity loss and the reuse loss:
\begin{align}
\mathcal{L} &= \mathcal{L}_{\text{sim}} + \alpha \cdot \max \left( 0, R_{\text{target}} - \mathcal{L}_{\text{reuse}} \right)
\label{eq:final-loss}
\end{align}
Here, $\alpha$ is a weighting hyperparameter that balances the trade-off between accuracy (controlled by $\mathcal{L}_{\text{sim}}$) and computational efficiency (encouraged by $\mathcal{L}_{\text{reuse}}$), and $R_{\text{target}}$ is the target reuse rate.
The $\max$ ensures that the penalty is applied only when the reuse rate falls below the target, thus incentivizing the model to meet $R_{\text{target}}$.
The combined loss function effectively prevents the model from trivially minimizing error through complete recomputation.
\subsection{Grouped Frame Training for Robustness}
\label{sec:grouped-frame-training}
%
Errors accumulate over time when computations are reused across multiple frames.
To ensure robustness, we model error accumulation during training by adopting a grouped frame training strategy.

\niparagraph{Modeling error accumulation.}
During training, we adjust our loss functions to account for grouped frames.
Specifically, we compute the losses over groups of frames by averaging the similarity loss $\mathcal{L}_{\text{sim}}$ and the reuse loss $\mathcal{L}_{\text{reuse}}$ across all frames in the group.
By considering the average losses over the group, the model learns to handle error accumulation effectively over sequences of frames, rather than optimizing for individual frames in isolation.
This approach ensures that the model’s performance and reuse behavior are optimized not just for single frames, but for sequences where errors may propagate due to computation reuse.

\niparagraph{Efficient grouping strategy.}
To balance modeling accuracy and training efficiency, we design training sequences that model error accumulation over a manageable number of frames.
\begin{figure}
    \centering
    \includegraphics[width=0.95\linewidth]{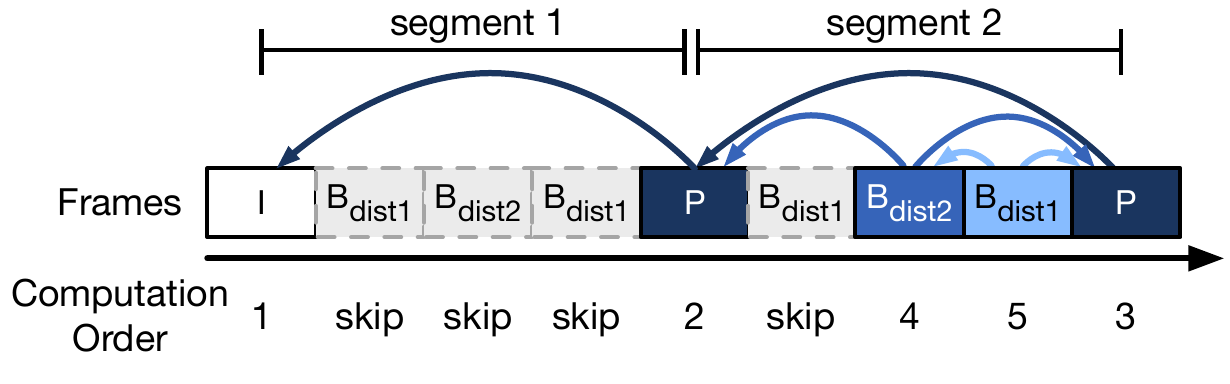}
    \caption{Illustration of our frame-grouping strategy during training. We treat each “I” or “P” frame as a segment boundary and form groups allowing the model to learn error accumulation.}
    \label{fig:grouping}
\end{figure}
We use I-frames and P-frames as segment boundaries, dividing frames into segments (e.g., Segment 1 and Segment 2 in Figure~\ref{fig:grouping}).
Frames within a segment can influence each other through computation reuse, while intermediate frames across segments remain independent, allowing for efficient modeling of longer sequences.

By constructing sequences like I-P-P, corresponding to frames 1, 5, and 9, we simulate error propagation over longer intervals without including intermediate frames (2–4 and 6–8).
Within the last segment, we also include frames with remaining B\textsubscript{dist2} and B\textsubscript{dist1} reference types from the previous segment, so that \reusevit is able to learn all reference types.

Empirical results indicate that grouping six frames in the pattern 1-5-9-13-11-12 strikes an effective balance between accuracy and training efficiency, allowing the model to account for error accumulation over significant temporal spans while managing computational costs.
During training, we also experimented with varying group sizes and found that 6 frame groupings gave us the best empirical tradeoff between training time and error-accumulation robustness.
Larger groupings occasionally made optimization less stable, while smaller ones did not fully capture longer-range drifts.

\section{Compaction Techniques in \dejavu} 
\label{sec:system-opt}
In this section, we present the compaction methods used to translate the FLOPs reduction achieved by \reusevit into tangible performance gains.
These methods address practical challenges in implementing \reusevit efficiently on GPU hardware, where parallelism is crucial to achieving high throughput.

\subsection{Layer-Wise Scheduling}
A key enabler of our compaction methods is layer-wise scheduling.
While the target problem differs, our scheduling scheme is inspired by FlexGen~\cite{flexgen}, a prior work that aims to deploy a large language model (LLM) serving on a single GPU.
FlexGen proposes grouping multiple batches for LLM inference and iterates the batches in a round-robin to execute their computations layer-by-layer.
That way, FlexGen can amortize the I/O costs across multiple batches, required for model parameters and KV cache.

On the other hand, \dejavu has different objectives since it aims to accelerate ViT encoders, which have smaller model sizes than LLMs and do not need KV cache.
\dejavu employs the layer-by-layer computation scheme to mitigate the memory bloating and GPU unfriendliness problems caused by \reusevit's computation reuse.
Instead of interleaving multiple batches as in FlexGen, \dejavu batches multiple frames together and processes the same layer across these frames in a layer-by-layer manner within the same segment.
%
Building upon this layer-wise scheduling, \dejavu employs Cached Memory Compaction and Sparse Computation Compaction to more efficiently utilize GPU memory and compute resources.
\subsection{Cached Memory Compaction}
\label{subsec:cached-memory-compaction}
A major overhead in computation reuse arises from storing cached activations, or intermediate outputs computed for reference frames, which subsequent frames rely on.

\niparagraph{Layer-wise memory compaction.}
By exploiting our layer-wise scheduling and the frame-referencing structure of our model (with P-frames acting as boundaries), we compact memory usage at each layer within a segment.
Frames other than P-frames do not affect future segments, so their intermediate activations can be discarded once their current segment is processed.
\begin{figure}
    \centering
    \includegraphics[width=\linewidth]{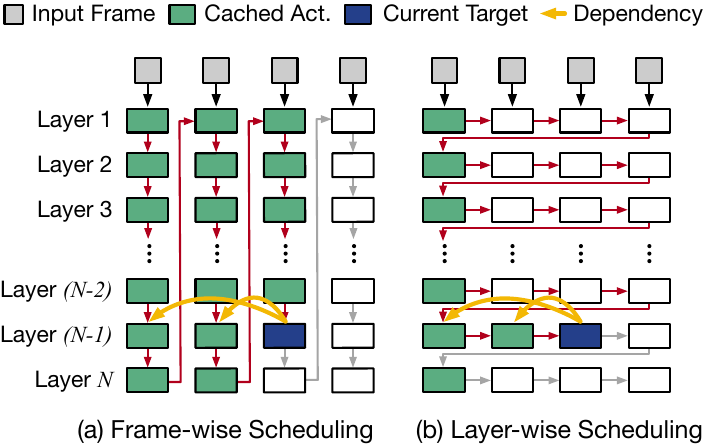}
    \caption{Comparison of (a) frame-wise scheduling and (b) layer-wise scheduling for caching intermediate activations.}
    \label{fig:layerwise}
\end{figure}

Figure~\ref{fig:layerwise} illustrates the key difference between the conventional frame-wise approach.
As illustrated in Figure~\ref{fig:layerwise}(a), in the conventional frame-wise approach, each frame runs through all layers sequentially.
Intermediate outputs must remain in memory until all computations for that frame are complete.
In contrast, in our layer-wise scheduling (Figure~\ref{fig:layerwise}(b)), the same layer is applied to all frames in the batch, and once a layer is complete for each frame, any unneeded activations are immediately freed.
This staggered approach substantially lowers memory overhead, enables larger batch sizes with high GPU utilization, and facilitates deployment of large-scale VideoLMs under limited memory budgets.
\subsection{Sparse Computation Compaction}
\reusevit's dynamic and often sparse computation patterns pose difficulties for GPUs, which are optimized for dense, regular workloads.
To address this, we use stream compaction~\cite{streamcompaction}, gathering active (i.e., non-reused) tokens into contiguous memory regions so sparse computations can be converted into dense forms better suited to GPU acceleration.

\begin{figure}
    \centering
    \includegraphics[width=\linewidth]{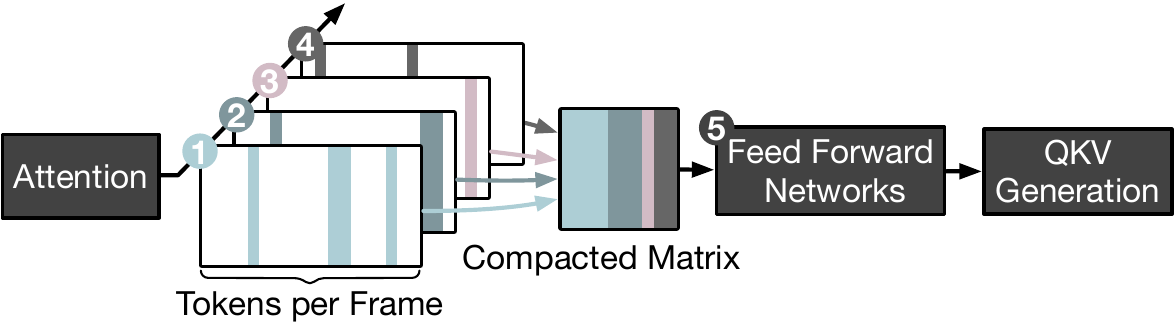}
    \caption{Sparse Computation Compaction.}
    \label{fig:compaction}
\end{figure}
\niparagraph{Layer-wise stream compaction.}
We implement GPU kernels for stream compaction across different frames within the same segment, as shown in Figure~\ref{fig:compaction}.
By accumulating active tokens from multiple frames before moving on to the next feed forward network, we form well-shaped matrices amenable to GPU acceleration.
Once the batched computation is complete, we scatter the results back into their original positions in each frame’s token sequence, preserving correctness for the next layer.
This gather-scatter strategy ensures that even when some frames have only a few active tokens, they can still benefit from dense GPU operations.

To minimize overhead further, we prioritize CUDA-based implementations for the gather-compute-scatter procedure, avoiding frequent CPU-GPU synchronization.
Such an approach not only speeds up the compaction process itself but also reduces latency spikes that might occur from excessive kernel launches or data transfers between CPU and GPU.


\section{Workflow of \dejavu}
\label{sec:workflow}
%
%

%
\subsection{Overview of Query Processing}
On a query, \dejavu returns cached embeddings if available; otherwise it generates them with \reusevit and stores the result.

\niparagraph{Embedding storage overhead.}
Storing frame-level VLP embeddings introduces minimal storage overhead.
For example, extracting embeddings at 2 FPS from a Full-HD H.264 video ($\sim$625KB/s) results in approximately 4KB/s of data (assuming 1024-dimensional FP16 embeddings at $\sim$2KB each).
This constitutes merely 0.64\% of the compressed video size, with further footprint reductions achievable via advanced floating-point compression techniques~\cite{chimp:vldb:2022, gorilla:vldb:2015, elf:vldb:2023}.
%
\subsection{Offline Preparation}
\label{subsec:offline-preparation}
Before handling real queries, \dejavu trains the decision and restoration modules for a given ViT backbone.
During training, we use the self-supervised objective, which combines a similarity loss (to match the original ViT outputs) and a reuse-based loss (to encourage high reuse rates).
The primary user-adjustable hyperparameter is the target reuse rate ($R_{\text{target}}$), defined in Equation~\ref{eq:final-loss}.
Training with different values of $R_{\text{target}}$ allows users to navigate various points on the accuracy-speed tradeoff curve.
Alternatively, one can specify a target cosine similarity, whereby the model learns to maximize the reuse rate while conforming to the accuracy threshold defined by that similarity value.

We also employ grouped-frame training to model how errors might accumulate.
This improves robustness, as the model learns to balance computational savings with accuracy preservation across different video types.
During training, Only the two lightweight modules are trained, while the pre-trained ViT backbone remains frozen and convergence typically occurs within an hour.
Such design significantly simplifies deployment by removing the need to store, transfer, or manage multiple versions of large model weights.
%
%
%
\subsection{Online Inference}
Once deployed, \dejavu employs layerwise scheduling and memory compaction to translate the FLOP savings from \reusevit into practical GPU throughput.
While \reusevit reduces the token-level computational load, effective scheduling and compaction routines ensure these savings materialize in actual GPU execution time, which is a key requirement for large-scale query systems.

To maximize GPU utilization, the system processes multiple videos in a single batch.
If only one long video is available, it is split into multiple, uniformly sized segments.
For each segment, four consecutive frames are collected via a small queue and fed into \reusevit, enabling each inference call to process four times the usual batch size in frames.
Batching multiple segments is crucial for efficiently utilizing GPU resources, especially when high reuse rates substantially reduce computation per video.

\niparagraph{Frame reordering.}
Frame reordering is handled within the model’s forward pass, which reconstructs results in the correct sequence.
Consequently, the outer framework does not need to manage any frame shuffling or realignment.
Although this staggered approach works well for offline analytics, where latency is less critical, it can also be adapted to streaming scenarios.
In a streaming setting, \dejavu buffers four frames before computation, introducing a queuing latency of about two seconds at 2 FPS.
For real-time applications with strict latency requirements, frame reordering can be disabled to minimize latency, albeit at the cost of lower reuse rates.

\niparagraph{Breaking error propagation.}
Although grouping frames largely confines errors, prolonged reuse can still amplify minor inaccuracies.
To counter this, \dejavu periodically inserts a fully recomputed I frame.
For instance, if the system resets every twentieth frame, it prevents indefinite error buildup without incurring excessive recomputation costs.
This method keeps the overhead below 5\% while maintaining accuracy over long sequences.


\section{Evaluation}
\label{sec:evaluation}
\begin{figure*}[t]
    \centering
    \includegraphics[width=0.95\linewidth]{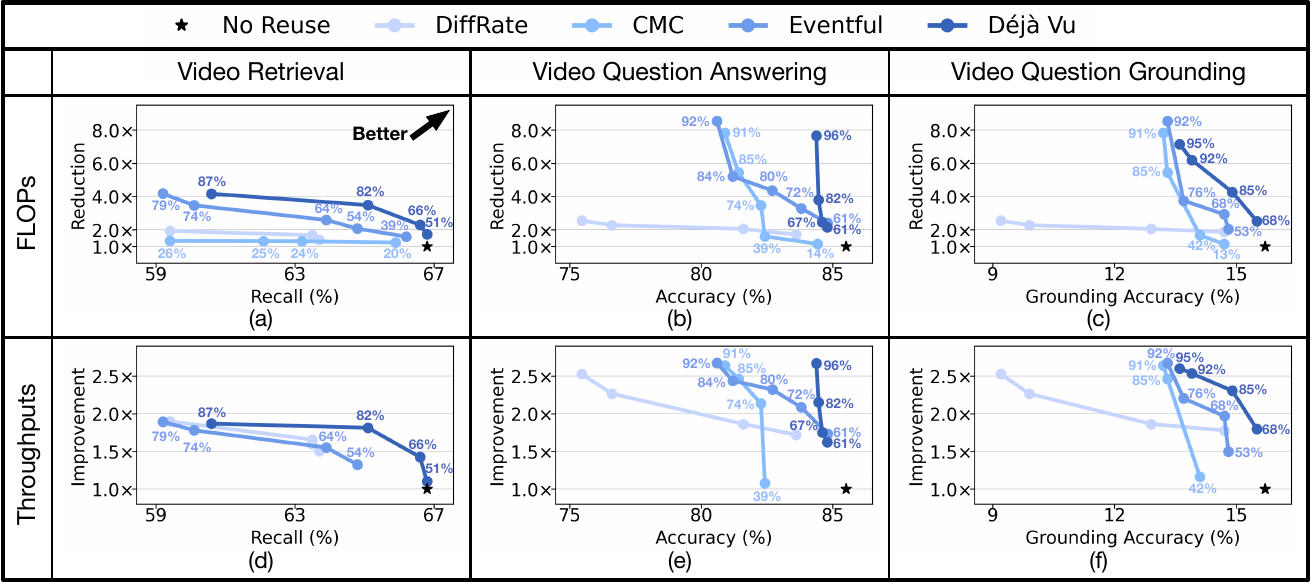}
    \caption{Tradeoff space of baselines and \dejavu. Y-axis for FLOPs reduction and throughput are normalized to original model with no reuse.}
    \label{fig:eval-tradeoff}
\end{figure*}
\subsection{Methodology}
\label{sec:methodology}
\niparagraph{End tasks.} 
To evaluate \dejavu's effectiveness, we select three distinct Video-Language Models (VideoLMs), each for a different task.
All models use CLIP~\cite{clip} (ViT) embeddings as their visual backbone.
We focus on:
\begin{itemize} [labelindent=0.3em,leftmargin=1.0em]
\item \textbf{Video retrieval.}
We use CLIP4Clip~\cite{clip4clip} on the MSR-VTT~\cite{msrvtt} dataset, which contains 10,000 text-annotated video clips of 10 to 30 seconds in length, and report top-5 recall.
\item \textbf{Video question answering.}
For video QA, we use FrozenBiLM~\cite{frozenbilm} on How2QA~\cite{how2qa}, which has 44,007 QA pairs on 60-second clips, and report multiple-choice accuracy.
\item \textbf{Video Question Grounding.}
For video question grounding, which requires localizing a temporal segment for an answer, we use TempCLIP~\cite{nextgqa:cvpr:2024} on NExT-GQA~\cite{nextgqa:cvpr:2024}, with videos averaging 45 seconds.
We report GQA@Accuracy, measuring correct answers with correct temporal overlap.
\end{itemize}
For a fair comparison, all methods sample frames at 2 FPS, aligning with common practice~\cite{rethinking:cvpr:2023, vivit:iccv:2021, stmixing:NeurIPS:2021, videollama:emnlp:2023, internvid:iclr:2024, clip4clip, frozenbilm, nextgqa:cvpr:2024}.
We use standard dataset splits for training and evaluation for all tasks.

\niparagraph{Baselines.}
As no prior methods directly accelerate VideoLMs, our baselines comprise conceptually similar temporal ViT acceleration approaches and a representative image-based method for contrast.
We evaluate three acceleration methods:
\begin{itemize} [labelindent=0.3em,leftmargin=1.0em]
\item \textbf{CMC.}
CMC~\cite{cmc} identifies tokens for reuse via a fixed mean squared error (MSE) threshold.
Originally designed for action recognition, it can leverage specialized video codec hardware for its similarity computations.
\item \textbf{Eventful Transformer.}
Eventful Transformer~\cite{eventful} employs a simple but rigid static policy, recomputing a fixed number of tokens per layer regardless of video content.
While simple and efficient, this non-adaptive strategy can cause error accumulation in dynamic scenes.
\item \textbf{DiffRate.}
DiffRate~\cite{diffrate} is an image-based baseline, adapted for VLP models, that combines token pruning and merging.
It automatically determines the accuracy-efficiency tradeoff through backpropagation.
Though originally evaluated on image classification, we adapted the policy for VLP models.
\end{itemize}
CMC and Eventful Transformer propose kernels to reuse computations in the attention layer.
In our experiments, these kernels did not yield speedups because the attention layers have relatively small computational costs.
Therefore, for both methods, we focus on their core reuse strategies without attention-layer optimizations.

\niparagraph{Implementation details.}
We run all experiments on a server with two 16-core Intel Xeon Gold 6226R CPUs (2.9 GHz), 192 GB of RAM, and an NVIDIA RTX 3090 GPU (24 GB of GDDR6 memory).
The software environment includes Ubuntu 24.04 as the operating system, CUDA 12.1, cuDNN 8.9.0, and PyTorch 2.1.0.
\subsection{Tradeoff between FLOPs and Accuracy}
Figure~\ref{fig:eval-tradeoff} (a)–(c) shows tradeoffs between FLOPs reduction for embedding generation (y-axis) and accuracy (x-axis) for three tasks: video retrieval, video QA, and video question grounding.
The x-axis shows the accuracy metric, and the star-shaped marker near the bottom right of each graph indicates the unapproximated model's accuracy.
The y-axis shows the achieved FLOPs reduction for embedding generation.
A point higher on the y-axis indicates greater FLOPs reduction at the same accuracy level, while a point further to the right indicates better accuracy.
Thus, methods near the top-right corner represent more favorable efficiency-accuracy tradeoffs.
Next to each marker, we show the computation reuse rate, except for DiffRate, which applies token pruning and merging instead.

Notably, techniques that exploit temporal redundancy (CMC, Eventful Transformer, and \dejavu) consistently achieve higher FLOPs reduction than the image-focused DiffRate, whose pruning/merging targets only spatial redundancy within individual frames.
Among these temporal-based baselines, \dejavu attains the best tradeoff, typically matching or exceeding each method’s accuracy at a lower computational cost.
\subsection{Tradeoff between Throughput and Accuracy}
Figures~\ref{fig:eval-tradeoff} (d)-(f) show throughput improvement for the embedding generation on the y-axis.
While \dejavu and DiffRate are measured on a real GPU system, CMC and Eventful Transformer lack GPU-optimized implementations, so we interpolate their throughput assuming our compaction techniques applied.
Here, we normalize throughput to that of the original model, and only configurations that yield an actual improvement are shown, and each marker can be matched to those in the FLOPs plots via the reuse rates.

Consistent with previous FLOPs-based results, \dejavu achieves the highest throughput-accuracy tradeoff.
For instance, \dejavu reaches speedups of up to 1.81$\times$ for video retrieval, 2.64$\times$ for video question answering, and 2.54$\times$ for video question grounding.
By contrast, Eventful Transformer, the next-best baseline, achieves speedups of 1.32$\times$, 2.08$\times$, and 2.20$\times$, respectively.

While baselines exploiting inter-frame redundancies remain competitive for VideoLM tasks, \dejavu's tailored optimizations achieve superior throughput-accuracy tradeoffs.
Even incremental improvements are crucial, for instance, a 1–2\% accuracy gain in leading video QA benchmarks typically requires several months of intensive research~\cite{frozenbilm,justask}, and modest speedups can significantly reduce costs and query latency in large-scale deployments.

Interestingly, DiffRate configurations become more competitive for throughput, as the high complexity of computation reuse can hinder its realization. However, \dejavu's careful optimizations still ensure higher overall speedups.
\subsection{Breakdown Analysis}
\label{subsec:breakdown-analysis}
At the same reuse rate, \dejavu has slightly lower performance than CMC or Eventful Transformer.
For instance, in Figure~\ref{fig:eval-tradeoff}~(e) when reuse is at 61\%, \dejavu shows a marginally lower throughput.
\begin{figure}
    \centering
    \includegraphics[width=1.0\linewidth]{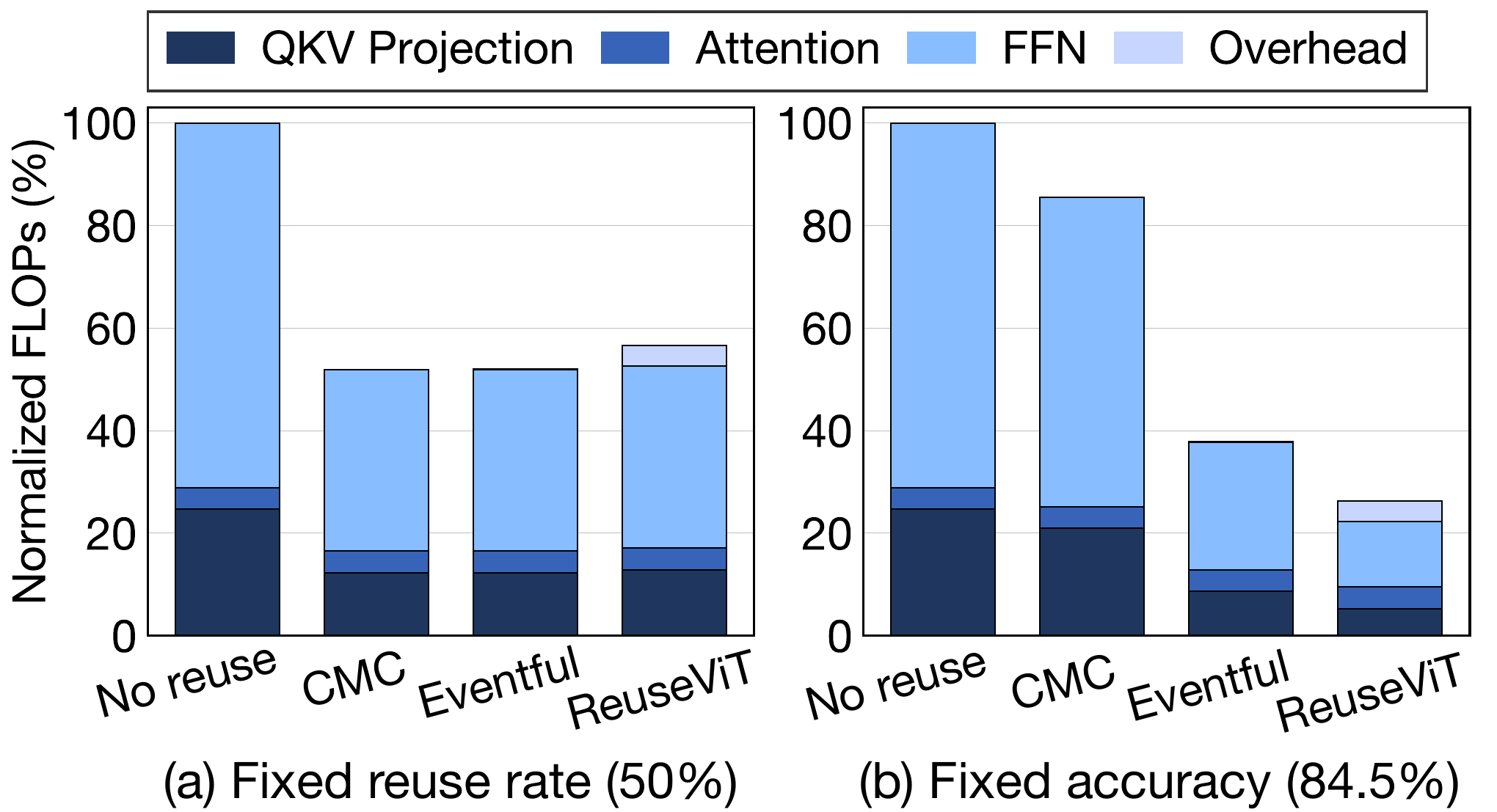}
    \caption{Breakdown of FLOPs on video retrieval. (a) FLOPs breakdown at fixed reuse rate. (b) Reuse rates comparison at equal accuracy.}
    \label{fig:eval-dejavu-flops-breakdown}
\end{figure}
Figure~\ref{fig:eval-dejavu-flops-breakdown}~(a) shows a FLOPs breakdown of these methods at an identical reuse rate on the video retrieval task.
In this scenario, the reuse rate is intentionally fixed across all methods to isolate and highlight algorithmic overhead.
The main additional overhead in \reusevit arises from the decision layers and restoration layers, which add about 4\% to normalized FLOPs.

However, Figure~\ref{fig:eval-dejavu-flops-breakdown}~(b) shows that at the same accuracy on video question answering, \dejavu achieves a significantly higher reuse rate.
Thus, at equivalent accuracy levels, \dejavu effectively compensates for the additional overhead by adaptively calibrating reused computations, enabling it to reuse more tokens without sacrificing accuracy.
Hence, while \dejavu incurs some overhead at the same reuse rate, its adaptive mechanisms enable higher reuse where accuracy must be maintained, which ultimately yields stronger end-to-end speedups when matching target accuracy levels.
\subsection{Memory Overhead}
\begin{figure}
        \centering
        \includegraphics[width=0.9\linewidth]{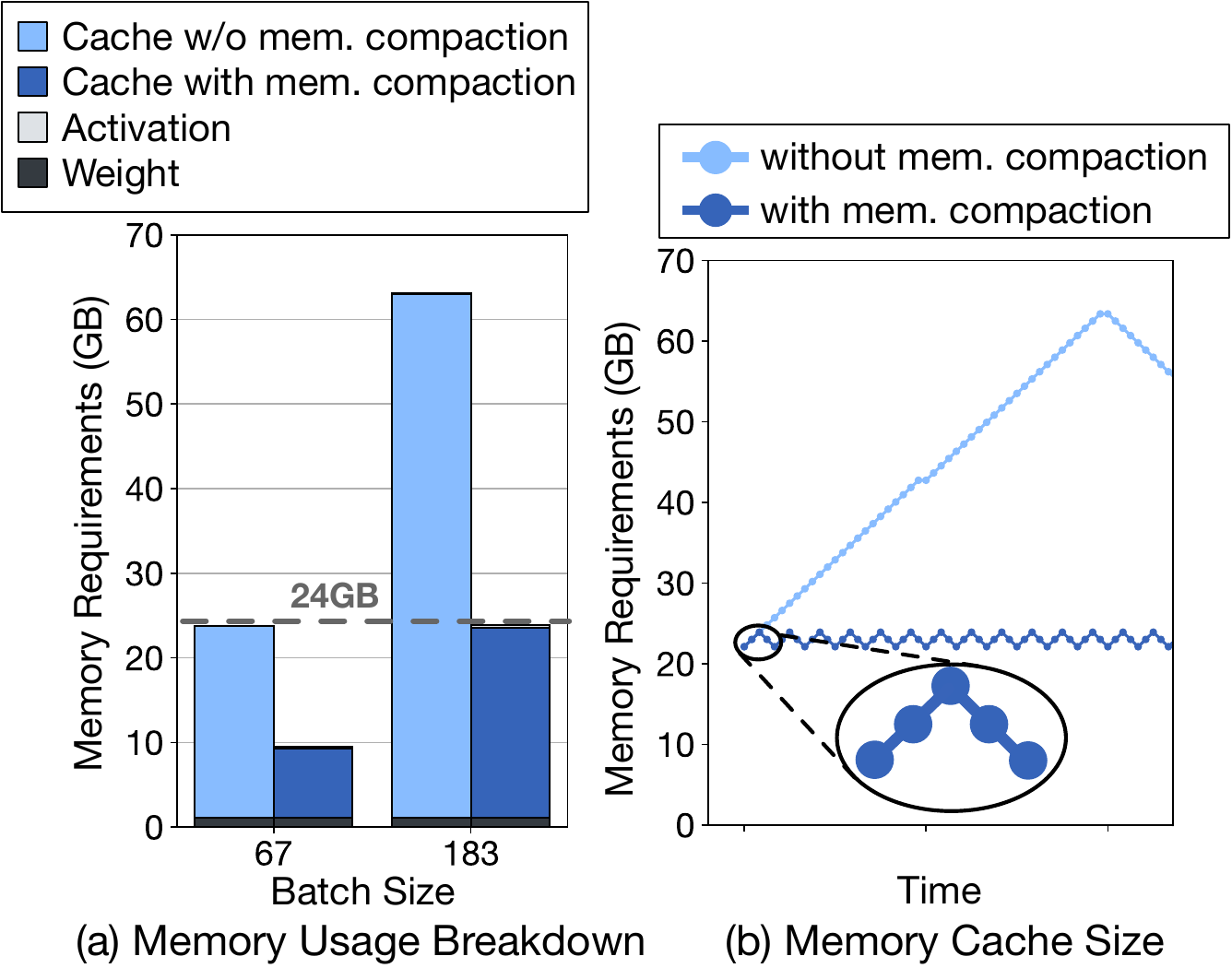}
        \caption{Comparison of peak memory usage with and without cached memory compaction. The dotted line shows memory capacity of RTX 3090 GPU.}
        \label{fig:tmp-memory-profile}
\end{figure}
Figure~\ref{fig:tmp-memory-profile}(a) compares the peak GPU memory usage of \reusevit with and without cached memory compaction, across different batch sizes.
The blue bars show memory usage when intermediate computations are cached without compaction, while the green bars include compaction.
Without compaction, the model processes only 67 batches before running out of memory, but compaction raises this limit to 183, improving GPU utilization and throughput.
By keeping memory usage within GPU limits, practitioners can deploy larger batch sizes or process multiple video streams concurrently, which in turn boosts overall system throughput.
This is particularly valuable in server-based or cloud environments where memory is a critical and often expensive resource.

Figure~\ref{fig:tmp-memory-profile}(b) illustrates how the cache size changes during inference.
Without compaction, the cache grows linearly until all layers are processed.
With compaction, cached memory for processed frames is freed after certain layers, creating a sawtooth pattern and significantly reducing peak memory usage.
\subsection{Ablation Study for Inference Speedup}
\begin{figure}[t]
    \centering
    \includegraphics[width=0.95\linewidth]{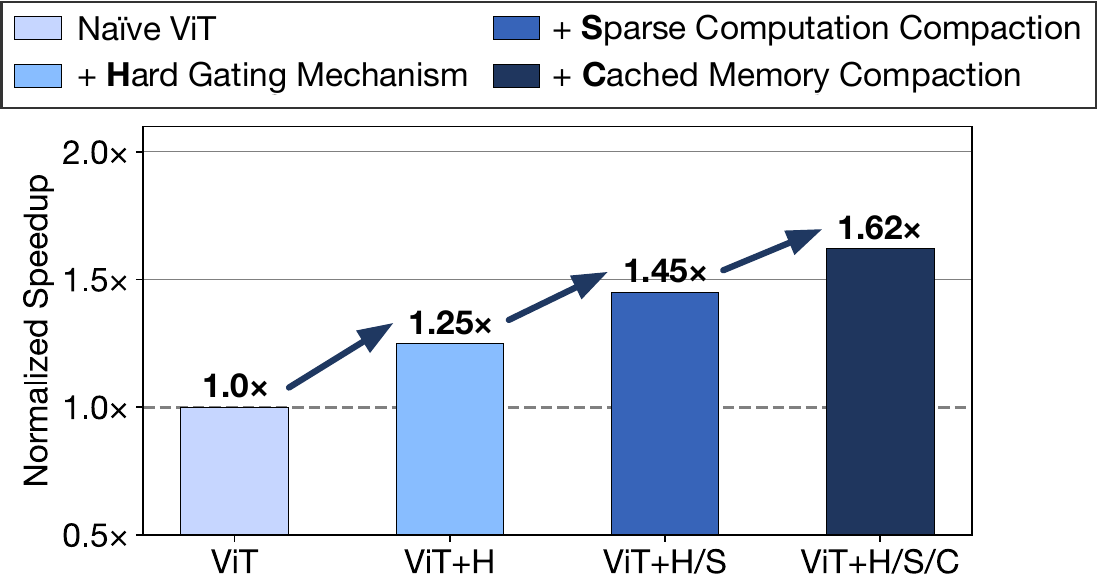}
    \caption{Ablation study of \dejavu's inference optimization techniques on the video QA task.}
    \label{fig:eval-ablation}
\end{figure}
We conduct an ablation study of inference speedup with a \reusevit configuration achieving a 61\% reuse rate on the video question answering task to show how each technique contributes to \dejavu's performance.
Figure~\ref{fig:eval-ablation} highlights the throughput gains from each step.
The first bar indicates reusing computation with hard gating alone yields a 1.25$\times$ speedup.
Layer-wise sparse computation compaction then increases the speedup to 1.45$\times$ by shaping operands to better utilize the GPU kernel.
Finally, layer-wise memory compaction enables larger batches, improving parallelization and raising the total speedup to 1.62$\times$.
These results confirm \dejavu's optimizations translate FLOPs savings into real throughput gains.
Each component offers complementary benefits, gating reduces compute, sparse-compaction improves GPU workload efficiency, and memory compaction allows larger batches, unlocking significantly more performance together than individually.
\subsection{Ablation Study for Design Choices}
\label{subsec:ablation-design-choice}
To evaluate the contributions of \reusevit's core components, we perform an ablation study on the NeXT-GQA dataset.
\begin{figure}[t]
    \centering
    \setlength{\fboxrule}{1pt}
    \includegraphics[width=\linewidth]{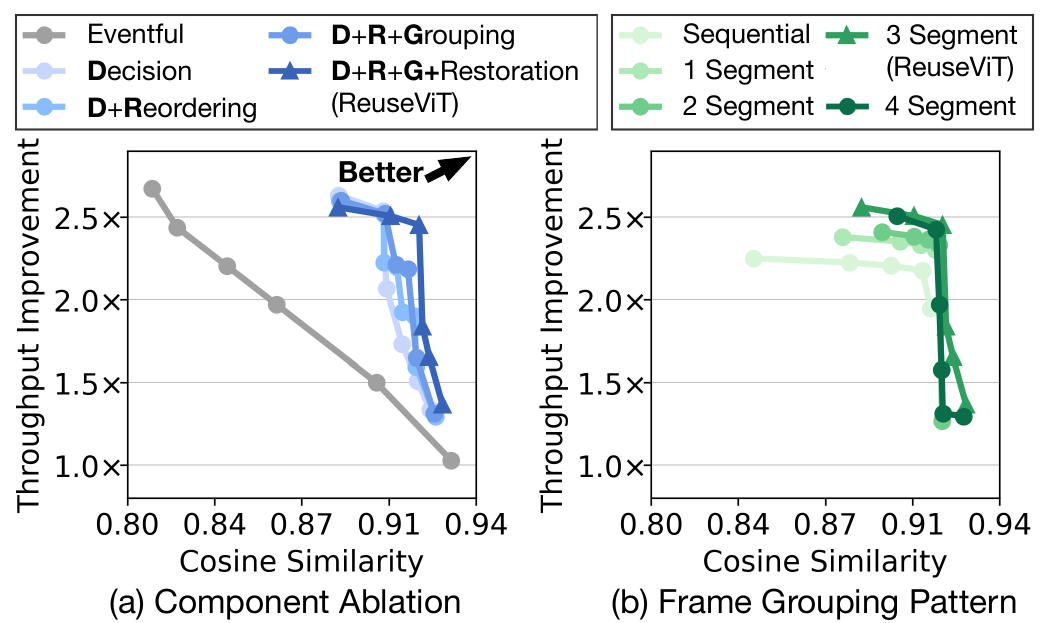}
    \caption{Ablation study of \reusevit's design choices on the NeXT-GQA dataset. The full \reusevit configuration is denoted by a triangular marker.}
    \label{fig:eval-ablation-design}
\end{figure}
Figure~\ref{fig:eval-ablation-design} assesses visual embedding quality, measured by cosine similarity to the original embedding without reuse, and throughput improvement normalized against the original ViT.
Higher values on both axes indicate better efficiency and accuracy.
Figure~\ref{fig:eval-ablation-design}(a) presents incremental component comparisons.
Starting from a configuration with only the adaptive decision layer, we sequentially incorporate frame reordering, grouped frame training, and finally, the restoration layer.
Notably, even the initial decision layer alone already provides substantial improvements over the static Eventful Transformer baseline, highlighting the significant advantage of adaptive reuse decisions.
Each subsequent addition further enhances the tradeoff between efficiency and accuracy.
Although the restoration layer introduces a slight computational overhead reducing peak throughput, it notably improves embedding quality at higher reuse rates.

Figure~\ref{fig:eval-ablation-design}(b) examines various frame grouping strategies.
Compared to a sequential processing baseline, all reordered groupings achieve improved tradeoffs.
Performance consistently increases with segment lengths up to three segments, while extending to four segments yields slightly inferior results.
Thus, a three-segment grouping strategy effectively balances computational efficiency and visual embedding quality.
\subsection{Adaptability to Video Content}
We compare the adaptability of Eventful Transformer and \dejavu on a video segments from How2QA, with both methods reaching 84.5\% accuracy.
\begin{figure}
    \centering
    \includegraphics[width=0.9\linewidth]{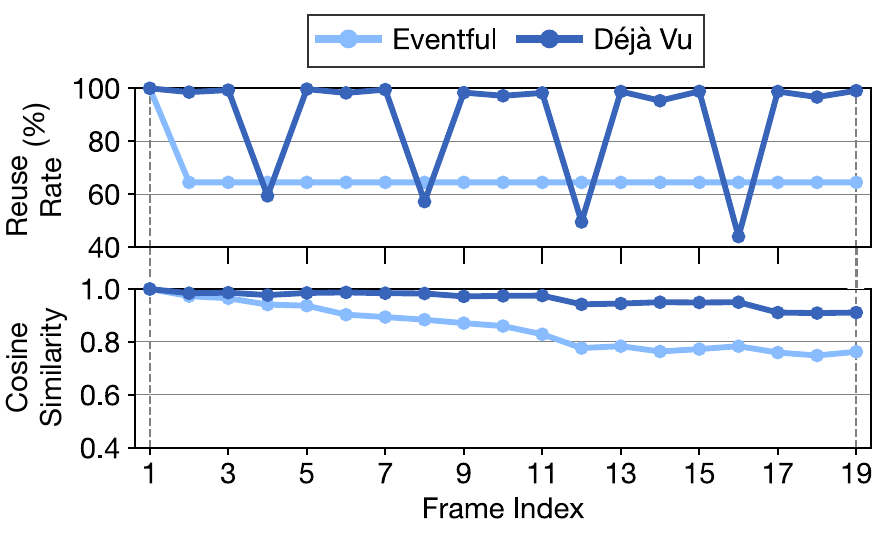}
    \caption{Comparison of \dejavu's adaptive reuse strategy versus Eventful Transformer's static strategy on a video segment from How2QA over time.}
    \label{fig:eval-adaptation}
\end{figure}
Figure~\ref{fig:eval-adaptation} depicts the fluctuations in computation reuse rates (top) and cosine similarities (bottom) over time.

Initially, both methods fully compute the first frame, then adopt their reuse strategies.
Eventful Transformer maintains a fixed reuse rate, leading to a more pronounced decline in cosine similarity.
In contrast, \dejavu adapts when error begins to accumulate, lowering the reuse rate to protect accuracy and raising it when frames are more similar.
Moreover, \dejavu learns to adjust reuse behavior based on reference frame types (I- or P-frames), further mitigating error accumulation.
By dynamically balancing reuse and recomputation, \dejavu preserves higher feature similarity over time and achieves a superior tradeoff between computational efficiency and model accuracy compared to methods with fixed reuse strategies.

\section{Limitations and Future Work}
\label{sec:limitation}
While \dejavu largely improves computational efficiency for video-language tasks, several limitations remain, motivating future work.

\niparagraph{Attention Layer Reuse.}
\dejavu currently targets only FFN and QKV layers. These layers dominate computational costs at standard resolutions, for instance, 257 tokens for ViT-L/14.
However, as resolution and token counts rise, such as 577 tokens (CLIP ViT-L/14 at 336px), 785 tokens (DINO ViT-B/8~\cite{dino:iccv:2021}), and 1370 tokens (DINOv2 ViT-G/14 at 518px~\cite{dinov2:tmlr:2024}), attention layers form an increasingly substantial fraction of total FLOPs, reaching up to 23.5\%.
Addressing this requires exploring attention reuse strategies or leveraging sparse attention kernels optimized for GPUs, like those proposed in Eventful Transformer.
Another approach involves specialized hardware accelerators, such as those introduced in CMC.
%

\niparagraph{Broader task generalization.}
This work evaluates \dejavu specifically on retrieval, question answering, and grounding tasks.
Adapting our learned reuse strategies to other video tasks, such as action recognition or object detection, would involve adjustments to task-specific model architectures and setups.
While this adaptation is conceptually straightforward, its practical exploration remains an open and promising direction for future research.

\niparagraph{Adaptive periodic refresh and online adaptation.}
Another limitation relates to the adaptive periodic refresh strategy within \dejavu. This strategy could benefit from additional refinement for improved handling of long-form video inputs and enhanced robustness across various video domains.
Furthermore, future research might explore online fine-tuning methods designed to dynamically adjust reuse policies during inference. Such adjustments would enhance adaptability and efficiency within evolving real-world scenarios.

\section{Related Work}
\label{sec:related}
\niparagraph{CNN-based pipelines for known classes.}
Early systems for known object categories often relied on trained CNNs or proxy models.
For instance, NoScope~\cite{noscope} developed lightweight cascades for static-camera queries.
Focus~\cite{focus:osdi:2018} and BlazeIt~\cite{blazeit} built approximate indexes that were refined by heavier models.
Other approaches included Tahoma's~\cite{tahoma:icde:2019} use of cascades with input transformations and MIRIS~\cite{miris:sigmod:2020} targeted multi-object tracking.
Some systems precomputed object tracks, like OTIF~\cite{otif:sigmod:2022}, or used detector ensembles, as seen in FiGO~\cite{figo:sigmod:2022}.
More recent contributions include DoveDB~\cite{dovedb:vldb:2023} unifying training and tracklet ingestion, Seiden~\cite{seiden:vldb:2023} leveraging high-accuracy models with sampling to avoid proxies, and InQuest~\cite{inquest:vldb:2023} paring cheap proxies with oracles for streaming analytics.
While cost-effective for predefined categories, these systems struggle with zero-shot or open-ended queries without retraining.
%

\niparagraph{Resource management and domain-specific analytics.}
Another line of work focuses on scheduling or adapting analytics pipelines at scale.
Scanner~\cite{scanner:tog:2018}, for example, exploits domain metadata for raw video decoding and dataflow.
VideoStorm~\cite{videostorm:nsdi:2017} and Chameleon~\cite{chameleon:sigcomm:2018} schedule model cascades for concurrent streams.
ODIN~\cite{odin:vldb:2020} monitors domain drift to retrain models for known classes under new conditions.
Further techniques involve adaptive query reordering by Hydro~\cite{hydro:2024}, using geospatial metadata to skip frames as in Spatialyze~\cite{spatialyze:vldb:2024}, or integrating caching to accelerate iterative queries, demonstrated by VIVA~\cite{viva:cidr:2022}.
%

\niparagraph{Open-vocabulary indexing and domain-agnostic exploration.}
To address fixed label sets, several systems enable querying arbitrary classes without retraining.
Panorama~\cite{panorama} learns a generic feature space.
TASTI~\cite{tasti:vldb:2021} reuses a learned semantic index for new labels, and Boggart~\cite{boggart:nsdi:2023} provides a model-agnostic index.
Other systems support interactive exploration, such as EVA~\cite{eva:sigmod:2022}.
VOCALExplore~\cite{vocalexplore:vldb:2023} trains detectors on-the-fly via active labeling, while SeeSaw~\cite{seesaw:pacmmod:2023} leverages CLIP embeddings for zero-shot retrieval.
LVS~\cite{lvs:ecv:2024} approximates embeddings from cached ones, and SketchQL~\cite{sketchql:pacmmod:2024} uses sketch-based queries.
Domain-agnostic storage solutions like VStore~\cite{vstore:eurosys:2019}, TASM~\cite{tasm}, and VSS~\cite{vss}, along with decoding-stage acceleration in CoVA~\cite{cova} and TVM~\cite{tvm:vldb:2023}, also enhance query speed.
%

\niparagraph{Modified architectures and acceleration for ViTs.}
Following ViT's~\cite{vit} success, many subsequent architectures aim to reduce computation using techniques such as local attention or convolutions~\cite{swintransformer,pvt,pvtv2,twins,edgevit,pit,levit}.
For video processing, temporal mixing layers are common additions~\cite{internvideo,vindlu,tevit,testa,videomae,videomae2,umt}, though no standard has emerged.
Standard ViTs are accelerated via token pruning (removing less important tokens)~\cite{dynamicvit,ats,evit,spvit,diffrate,pumer} or token merging (combining similar tokens)~\cite{tome,tokenpooling,diffrate,pumer,vidtome,vid-tldr}.
Pruning techniques can be fixed-rate~\cite{dynamicvit} or adaptive~\cite{ats}, sometimes combined with merging strategies~\cite{evit,spvit}.
Merging approaches, exemplified by ToMe~\cite{tome} and TCFormer~\cite{tokenpooling}, downsample tokens, with recent extensions to temporal attention~\cite{vidtome,vid-tldr}.
Some methods integrate both pruning and merging~\cite{diffrate,pumer} or propose specialized hardware~\cite{leopard,spatten,heatvit,fact,adaptiv:micro:2024}.
%

\niparagraph{Computation reuse and decision making.}
Reusing computations across frames offers gains beyond intra-frame redundancy.
Eventful Transformer~\cite{eventful} and CMC~\cite{cmc} use manual cross-frame reuse decisions.
In contrast, \dejavu employs a fully learnable mechanism for improved efficiency.
Prior work on temporal redundancy in CNNs includes incremental activation updates~\cite{eva2,va-red2}, use of motion vectors~\cite{vrdann}, delta updates~\cite{deltacnn,motiondeltacnn}, or discarding redundant frames and regions~\cite{dlta,respire,reducto,liteeval,activation-sparsity}.
Several studies also learn reuse decisions end-to-end.
These include methods for dynamic fusion~\cite{adafuse} or layer skipping~\cite{blockdrop,skipnet,dynamicconv}.
While these predominantly target CNNs, \dejavu adapts this concept to VideoLMs by integrating a learnable reuse mechanism in ViTs for significant speedups while preserving accuracy.
\section{Conclusion}
\label{sec:conclusion}
This paper addresses the computational challenges posed by ViT in the context of large-scale video analytics exploiting VideoLMs.
We introduce \dejavu, a novel algorithm-system co-designed solution, recognizing the untapped potential of exploiting a learning approach to enable inter-frame computation reuse.
Central to \dejavu is the \reusevit model, innovatively adapted for VideoLM tasks.
This model is specifically designed for enabling computation reuse opportunities in ViT inferencing by judiciously determining the reuse targets and restoring the reuse-introduced errors automatically.
We also build a \reusevit-based \dejavu system, which effectively utilizes GPUs for high throughput ViT inferencing for VideoLM queries.
Our evaluation shows that \dejavu achieves a substantial performance boost while marginally compromising the accuracy of VideoLM tasks. These findings suggest that this work represents a practical initial step toward actualizing the full potential of VideoLMs for video-language query systems.

%
%

\begin{acks}
This work was supported by the Institute of Information \& Communications Technology Planning \& Evaluation (IITP) grant funded by the Korea government (MSIT), through the Information Technology Research Center (ITRC) program (No. IITP-2025-RS-2020-II201795), the Development of ML compiler framework for on-device AI (No. RS-2024-00459797), and the Graduate School of Artificial Intelligence Semiconductor program (No. IITP-2025-RS-2023-00256472).
\end{acks}


\bibliographystyle{ACM-Reference-Format}
\bibliography{paper}

\end{document}